%Paper: hep-ph/9512392
%From: Simon Sirca <Simon.Sirca@ijs.si>
%Date: Thu, 21 Dec 95 12:43:14 MET

\font\mittel=cmbx10 scaled\magstep2
\font\bmath=cmmib10 scaled\magstep1
\magnification=\magstephalf
\vsize=24.181 true cm \hsize=16.425 true cm
\font\script=cmr8

\def\shiftdown#1{#1\llap{\lower.04ex\hbox{#1}}}

\def\np{$n_\pi$}
\newread\epsffilein    % file to \read
\newif\ifepsffileok    % continue looking for the bounding box?
\newif\ifepsfbbfound   % success?
\newif\ifepsfverbose   % report what you're making?
\newdimen\epsfxsize    % horizontal size after scaling
\newdimen\epsfysize    % vertical size after scaling
\newdimen\epsftsize    % horizontal size before scaling
\newdimen\epsfrsize    % vertical size before scaling
\newdimen\epsftmp      % register for arithmetic manipulation
\newdimen\pspoints     % conversion factor
\pspoints=1bp          % Adobe points are `big'
\epsfxsize=0pt         % Default value, means `use natural size'
\epsfysize=0pt         % ditto
\def\epsfbox#1{\global\def\epsfllx{72}\global\def\epsflly{72}%
   \global\def\epsfurx{540}\global\def\epsfury{720}%
   \def\lbracket{[}\def\testit{#1}\ifx\testit\lbracket
   \let\next=\epsfgetlitbb\else\let\next=\epsfnormal\fi\next{#1}}%
\def\epsfgetlitbb#1#2 #3 #4 #5]#6{\epsfgrab #2 #3 #4 #5 .\\%
   \epsfsetgraph{#6}}%
\def\epsfnormal#1{\epsfgetbb{#1}\epsfsetgraph{#1}}%
\def\epsfgetbb#1{%
\openin\epsffilein=#1
\ifeof\epsffilein\errmessage{I couldn't open #1, will ignore it}\else
   {\epsffileoktrue \chardef\other=12
    \def\do##1{\catcode`##1=\other}\dospecials \catcode`\ =10
    \loop
       \read\epsffilein to \epsffileline
       \ifeof\epsffilein\epsffileokfalse\else
          \expandafter\epsfaux\epsffileline:. \\%
       \fi
   \ifepsffileok\repeat
   \ifepsfbbfound\else
    \ifepsfverbose\message{No bounding box comment in #1; using defaults}\fi\fi
   }\closein\epsffilein\fi}%
\def\epsfsetgraph#1{%
   \epsfrsize=\epsfury\pspoints
   \advance\epsfrsize by-\epsflly\pspoints
   \epsftsize=\epsfurx\pspoints
   \advance\epsftsize by-\epsfllx\pspoints
   \epsfxsize\epsfsize\epsftsize\epsfrsize
   \ifnum\epsfxsize=0 \ifnum\epsfysize=0
      \epsfxsize=\epsftsize \epsfysize=\epsfrsize
     \else\epsftmp=\epsftsize \divide\epsftmp\epsfrsize
       \epsfxsize=\epsfysize \multiply\epsfxsize\epsftmp
       \multiply\epsftmp\epsfrsize \advance\epsftsize-\epsftmp
       \epsftmp=\epsfysize
       \loop \advance\epsftsize\epsftsize \divide\epsftmp 2
       \ifnum\epsftmp>0
          \ifnum\epsftsize<\epsfrsize\else
             \advance\epsftsize-\epsfrsize \advance\epsfxsize\epsftmp \fi
       \repeat
     \fi
   \else\epsftmp=\epsfrsize \divide\epsftmp\epsftsize
     \epsfysize=\epsfxsize \multiply\epsfysize\epsftmp
     \multiply\epsftmp\epsftsize \advance\epsfrsize-\epsftmp
     \epsftmp=\epsfxsize
     \loop \advance\epsfrsize\epsfrsize \divide\epsftmp 2
     \ifnum\epsftmp>0
        \ifnum\epsfrsize<\epsftsize\else
           \advance\epsfrsize-\epsftsize \advance\epsfysize\epsftmp \fi
     \repeat
   \fi
   \ifepsfverbose\message{#1: width=\the\epsfxsize, height=\the\epsfysize}\fi
   \epsftmp=10\epsfxsize \divide\epsftmp\pspoints
   \vbox to\epsfysize{\vfil\hbox to\epsfxsize{%
      \includegraphics{#1}%
      \hfil}}%
\epsfxsize=0pt\epsfysize=0pt}%
{\catcode`\%=12 \global\let\epsfpercent=%\global\def\epsfbblit{%BoundingBox}}%
\long\def\epsfaux#1#2:#3\\{\ifx#1\epsfpercent
   \def\testit{#2}\ifx\testit\epsfbblit
      \epsfgrab #3 . . . \\%
      \epsffileokfalse
      \global\epsfbbfoundtrue
   \fi\else\ifx#1\par\else\epsffileokfalse\fi\fi}%
\def\epsfgrab #1 #2 #3 #4 #5\\{%
   \global\def\epsfllx{#1}\ifx\epsfllx\empty
      \epsfgrab #2 #3 #4 #5 .\\\else
   \global\def\epsflly{#2}%
   \global\def\epsfurx{#3}\global\def\epsfury{#4}\fi}%
\def\epsfsize#1#2{\epsfxsize}

\def\d{{\rm d}}
\def\bra{\langle}
\def\ket{\rangle}
\def\a{a^{\phantom{\dagger}}}
\def\ac{a^{\dagger}}

\vbox{\vskip 3.1 true cm}
\leftline{\hskip 2.8 truecm
{\bf \mittel  Counting pions in the nucleon
} }
\vskip 0.75 truecm
\leftline{\hskip 2.8 truecm  M. Fiolhais$^1$, B. Golli$^{2,4}$,
M. Rosina$^{3,4}$ and S. \v Sirca$^4$ }
\par \bigskip
\leftline{\hskip 2.8 truecm {\sl $^1$ Departamento de F\'{\i}sica,
Universidade de Coimbra,
 P-3000 Coimbra, Portugal} }
\par \smallskip
\leftline{\hskip 2.8 truecm {\sl $^2$ Faculty of Education,
University of Ljubljana,
SI-61000 Ljubljana, Slovenia} }
\par \smallskip
\leftline{\hskip 2.8 truecm {\sl $^3$ Faculty of Mathematics and Physics,
University of Ljubljana,} } \par \smallskip
\leftline{\hskip 3.8 truecm
 {\sl  P.O.B. 64, SI-61111 Ljubljana, Slovenia} }
\par \smallskip
\leftline{\hskip 2.8 truecm {\sl $^4$ Jo\v zef Stefan Institute,
  P.O.B. 100, SI-61111 Ljubljana, Slovenia} }

\vskip 1 true cm

\leftline{\hskip 2.8 truecm  {\bf ABSTRACT} } \medskip
\noindent
The number of pions in the nucleon is a theoretical concept and its
value is model dependent. It is, however, a useful tool to understand
the pionic contribution to several nucleon observables
such as electric charge radii, magnetic moments,
axial charge, pion-nucleon coupling constant,
amplitudes for electroexcitation of baryon resonances
and electroproduction of pions above such resonances.

\bigskip \leftline{\hskip 2.8 truecm  {\bf KEYWORDS} } \medskip
\noindent
Pion cloud, static properties of the nucleon, electroexcitation of
$\Delta$, electroproduction of pions, linear sigma model, chiral
chromodielectric model, cloudy bag model
\bigskip \medskip

\leftline{\hskip 2.8 truecm  {\bf INTRODUCTION} } \medskip
\noindent

Many nucleon properties can be rather successfully explained by assuming
a cloud of pions surrounding the quark core.
We shall discuss here what information on the pion cloud we can extract
from {\it static  observables} of the nucleon as well as from
{\it dynamic observables} such as the amplitudes for electroexcitation
of baryon resonances. We shall particularly emphasize the role
of the number of pions (\np) as a tool to explore the relative
pion contribution to the observables and the model (in)dependence
of the results.
We shall try to distinguish two effects: the one in which the pionic
contribution is proportional to the number of pions, and the other where
it is proportional to the specific contribution of each pion which
depends on its radial profile.

Next, it is an important question whether $n_\pi$ is also a
useful phenomenological concept counting pions as partons
eagerly waiting to be kicked out in the process
$e N  \to e' X \pi$ (G\"uttner, Chanfray, Pirner and Povh, 1984).
The theoretical number of pions does not have
to coincide with the number of partons in an impulse
approximation, but it is
more ``natural" if it does. By ``natural" we mean that such a model
has a good chance to make realistic prediction without new ad hoc
parameters. An ``unnatural" or effective model, however, has to introduce
effective charges and effective currents in order
to relate the theoretical number of pions to the properties of the
electroproduction process.

Finally, \np \ is a useful mathematical-physical tool to determine
the regime of the quark-pion coupling: we have shown
(\v Cibej, Fiolhais, Golli and Rosina, 1992) that
for $n_\pi < 0.3 $ the solution is in the weak-coupling (perturbative)
regime, while for $n_\pi>3 $  it is in the strong-coupling regime.
In many models the best fit to experimental observables requires
$n_\pi \sim 1 $ so that none of the extremes applies.
This is an important warning to
assess the quality of some approximate schemes which unjustly assume either
perturbative regime or strong coupling regime. For example,
the semiclassical or cranking projection of angular momentum and isospin
in some formalisms is valid only in the strong coupling regime;
in the CBM, the first order perturbation is justified only for
$R\geq 1\,\hbox{fm}$.

The definition of the number of pions is somewhat arbitrary.
One possibility is to count the ``free pions" $\sum_{tklm} \ac_{tlm}(k)
\a_{tlm}(k)$ defined by the creation
and annihilation operators in the plane wave or spherical wave
expansion of the pion field operator:

$$\hat\pi_t({\bf r}) = \sum_{klm}\sqrt{{2\over\pi}}\,j_l(kr)
Y_{lm}(\hat{\bf r})
 {1\over \sqrt{2\omega_k}} [\a_{tlm}(k) + (-1)^{t+m}\ac_{-tl-m}(k)]$$
and to define observables as normal products with respect to the free vacuum
$|0\ket$,
$$ a_{tlm}(k) \vert 0 \ket = 0 .$$
Here $\omega_k^2=k^2+m_{\pi}^2$ and $\sum_k$ corresponds to $\int k^2dk$.

 Instead of ``free pions" we could have counted ``distorted pions"
$$\tilde{n}_\pi
= \sum_i {\tilde{a}^\dagger}_i {\tilde{a}^{\phantom{\dagger}}}_i $$
in a more general canonical basis defined by
$$\tilde{a}^\dagger_i = \sum_{tklm} x_{i,tklm} \ac_{tlm}(k) +
                                   y_{i,tklm} \a_{tlm}(k).    $$
The usefulness of such a Bogoliubov transformation has been explored
very little. We found in a variational calculation of the projected
hedgehog nucleon in a schematic model
that the additional variation of $x$ and $y$ coefficients
lowered the ground state energy at most by a few percent.
It would
be confusing if different authors were comparing the number of pions
in different Bogoliubov basis; therefore we advocate to compare
the number of free pions.

We should emphasize that a unitary transformation to ``distorted pions"
in which $\ac$ is only a superposition of $\ac$s without $a$s
($y=0$) does not change the number of pions, therefore it is the same
if we count ``plane wave pions" or ``spherical wave pions" or pions
with any radial profile.

\goodbreak

\bigskip
\medskip
\goodbreak
\noindent
\leftline{\hskip 2.8 truecm  {\bf SOME POPULAR MODELS FOR THE PION CLOUD}}
\medskip
\nobreak\noindent

We shall analyze the results of three different models, defined by
the corresponding Hamiltonians.

(i) Linear sigma model (LSM) (Birse, 1990; and references therein)
$$\eqalign{ H & = \int d^3x\, \biggl[{1\over 2}
  [\vec{P}^2_{\pi} + \nabla\vec{\pi}\cdot\nabla\vec{\pi}
  + P^2_{\sigma} + \nabla\sigma\cdot\nabla\sigma] \cr
  & + \bar\Psi[-\hbox{i} \hbox{\bmath\char13}\cdot\nabla +
     g(\sigma + \hbox{i} \gamma_5 \vec{\tau}\cdot\vec{\pi})]\Psi
  +{\lambda^2\over 4}(\sigma^2+\vec{\pi}^2-\nu^2)^2
  -f_\pi m_\pi^2 \sigma   \biggr] \>{,} \cr}  $$

(ii) Chiral chromodielectric model (CDM) (Birse, 1990)
$$\eqalign{ H & = \int d^3x \biggl\{{1\over 2}
  [\vec{P}^2_{\pi} + \nabla\vec{\pi}\cdot\nabla\vec{\pi}
  + P^2_{\sigma} + \nabla\sigma\cdot\nabla\sigma +
    P^2_{\chi} + \nabla\chi\cdot\nabla\chi] \cr
  & + \bar\Psi\biggl[-\hbox{i} \hbox{\bmath\char13}\cdot\nabla +
     {g \over \chi} (\sigma + \hbox{i} \gamma_5
     \vec{\tau}\cdot\vec{\pi})\biggr]\Psi
  +{\lambda^2\over 4}(\sigma^2+\vec{\pi}^2-\nu^2)^2
  -f_\pi m_\pi^2 \sigma    + {1 \over 2} M_\chi^2 \chi^2 \biggr\} \>{,}\cr}  $$

\hfill\eject

(iii) Cloudy bag model (CBM) (Thomas, 1983)

$$ H  = \int d^3x \biggl[{1\over 2}
  [\vec{P}^2_{\pi} + \nabla\vec{\pi}\cdot\nabla\vec{\pi}
  + m_\pi^2 \vec{\pi}^2 ]
  +{{\rm i} \over 2f_\pi} \bar\Psi \gamma_5
    \vec{\tau}\cdot\vec{\pi} \Psi \delta(x-R)
  - \hbox{i} \bar\Psi\hbox{\bmath\char13}\cdot\nabla\Psi\Theta(R-x)\biggr]\>{.}
        $$

In all these expressions, $\Psi$ stands for the quark field, $\vec \pi$ and
$\sigma$ for the chiral meson fields, and $\chi$ for the scalar-isoscalar
chiral
singlet field which (dynamically) generates  confinement in the CDM. The model
parameters are: the pion mass $m_\pi=0.14$ GeV and the pion decay constant
$f_\pi=0.093$ GeV which appear in the three models (these parameters also enter
in the mexican-hat potential for the chiral fields in LSM and CDM); the sigma
mass $m_\sigma=1.2$ GeV which  enters in the mexican-hat in LSM and CDM. In
the LSM there remains just one (dimensionless) parameter $g$. In the CDM
there are
two free parameters, the coupling constant $g$ (with dimension of energy)
and the $\chi$ mass $M_\chi$. However, in practice   the results are
sensitive mainly to one combination, $G=\sqrt{gM_\chi}$.
Finally, in the CBM the free parameter is the bag radius $R$.

The purpose of comparing the three models is
to generate situations with a wide range of $n_\pi$ and
to explore model dependence of our conclusions.

The first two models offer only a small range of $n_\pi$. In the LSM, for
small $n_\pi$ ($n_\pi \ll 1 $) the system of three quarks is not bound, and for
a large one ($n_\pi \gg 1 $) the vacuum
gets unstable against baryon-antibaryon creation.
Reasonable values of observables are obtained in the range
$4.5<g<5.5$ and $n_\pi \sim 1$.
In the CDM the system of three
quarks is always confined by the $\chi$ field and therefore the sigma and the
pion fields are much suppressed with respect to their values in the LSM.
Correspondingly, $n_\pi$ is smaller in comparison with its  typical values
in the LSM. Even for very high coupling constant $G$, the selfconsistent
field $\chi$ is such that  a strong effective coupling between
quarks and pions is never achieved.
Therefore the number of pions always remains small ($n_\pi \sim 0.2$).
The physical range of the coupling constant is
$0.175\,\hbox{GeV}<G<0.205\,\hbox{GeV}$.
The cloudy bag is a suitable model to offer a wide range of $n_\pi$
since the quarks are confined by a bag and pions are not indispensable
for binding. However, the number of pions is very sensitive to changes of $R$
so that, by varying $R$ in a wide range we can obtain $n_\pi$ in a wide range
too. In the presentation of our results,
we consider unrealistic values of $n_\pi$
just for the purpose of studying the $n_\pi$-dependence of the
observables.

{\epsfxsize11.863cm$$\epsfbox[90 490 520 750]{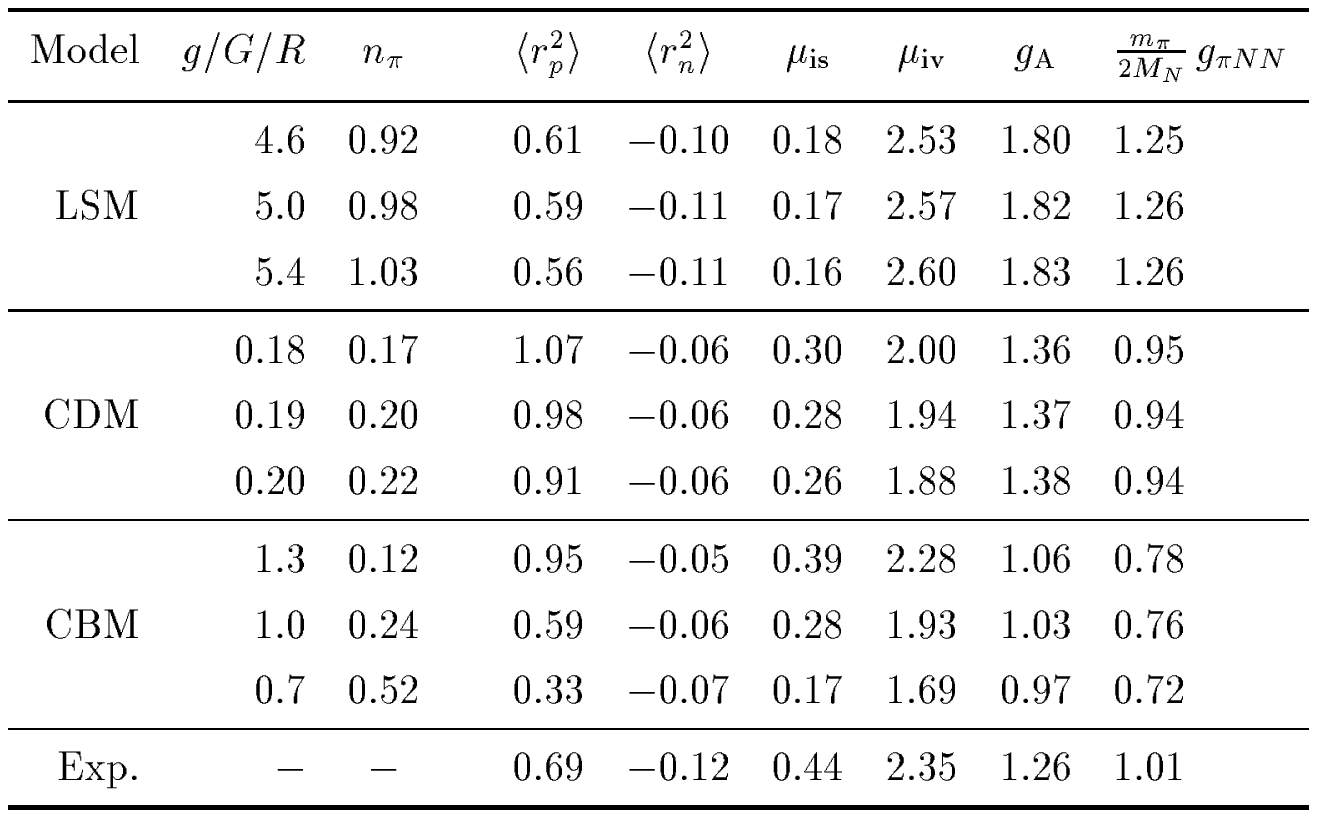}$$
\noindent Table 1: Static quantities of the nucleon shown in dependence of the
coupling constant $g$ (in LSM), $G$ [GeV] (in CDM)
and bag radius $R$ [fm] (in CBM) and of the number of pions.
Charge radii are in units of fm$^2$, magnetic moments in n.m.}

\goodbreak
\noindent
\leftline{\hskip 2.8 truecm  {\bf HOW STATIC PROPERTIES OF THE NUCLEON}}

\leftline{\hskip 2.8 truecm  {\bf DEPEND ON THE NUMBER OF PIONS}}
\medskip
\nobreak\noindent

We calculated nucleon properties in all three models using
a projected hedgehog wavefunction (\v Cibej, Fiolhais, Golli and Rosina, 1992)
 for the three valence quarks and the
meson (and chromodielectric) fields. The fields were treated
quantum mechanically as coherent states. All radial profiles were determined
variationally, with variation after projection, subject to virial constraints
of the type $\bra\Psi|[H,P({\bf r})]|\Psi\ket = 0$ (Amoreira, Fiolhais,
Golli and Rosina, 1995). Using such a constraint guarantees
the proper asymptotic behaviour of the pion field as well as some
other consistency relations. The calculated static
properties are shown in Table 1.

All figures show the observables as a function of \np.
Such diagrams were  not shown before and we believe that they are
very instructive. In order to recognize the respective coupling
strength in each model which corresponds to a given \np, the
``vocabulary" is given in Fig.~1.

{\epsfxsize13.7cm$$\epsfbox[30 590 520 750]{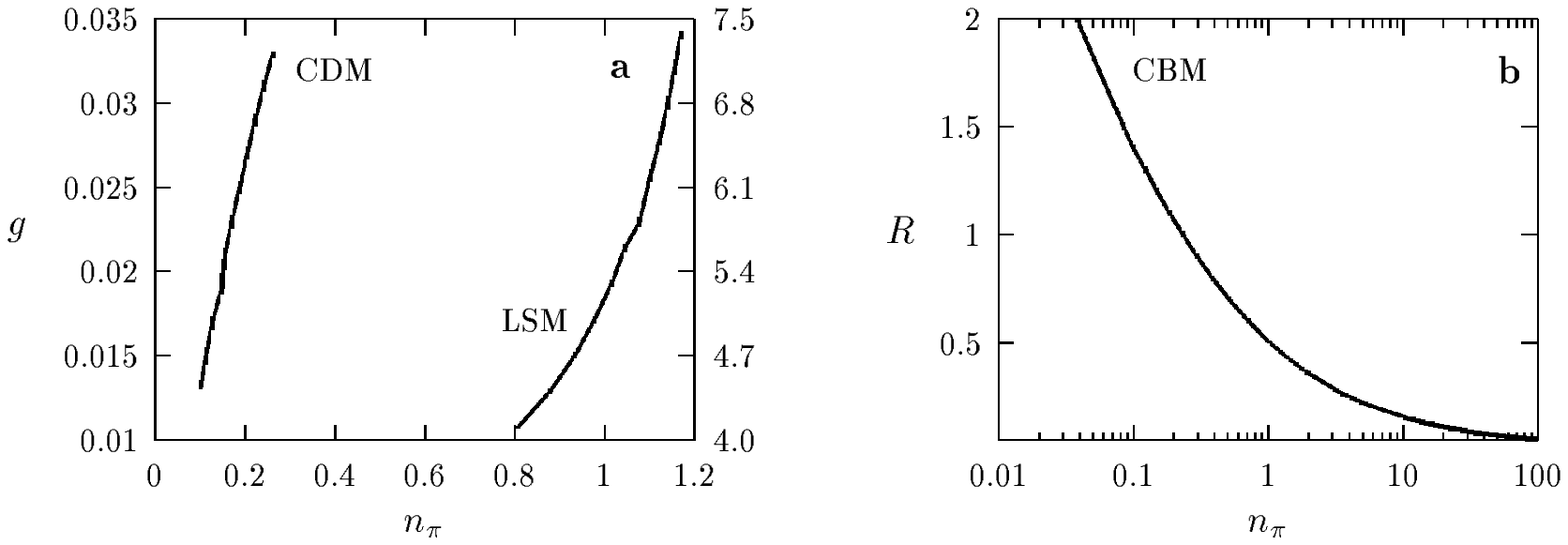}$$
\noindent Figure 1: Coupling constants $g$ [MeV] in CDM
(with $M_\chi$ fixed to $1.4\,\hbox{GeV}$)
and $g$ [dimensionless] in LSM ({\bf a}) and
bag radius $R$ [fm] in CBM ({\bf b}) as a function of $n_\pi$.}

\bigskip
\noindent
{\bf Nucleon--Delta mass difference}
\smallskip\noindent
In all three models the $\Delta-N$ mass splitting is mostly due to pions.
For low \np\ the mass splitting increases with the number of pions because
nucleon and $\Delta$ differ in the kinetic energy of pions but not in the
interaction energy of pions with the quark source;  with increasing
coupling constant the number of pions \np\ grows and also
the kinetic energy per pion grows (since all radial profiles shrink).
For high \np\ the mass splitting decreases because both states
belong to a rotational band and the moment of inertia increases with \np.
There is a peak between \np\ = 1  and  10 (Fig.~2).
It is favourable to have
\np\ around 1 in order to avoid exceedingly large additional terms
such as chromomagnetic repulsion.

{\epsfxsize9.125cm$$\epsfbox[80 480 470 700]{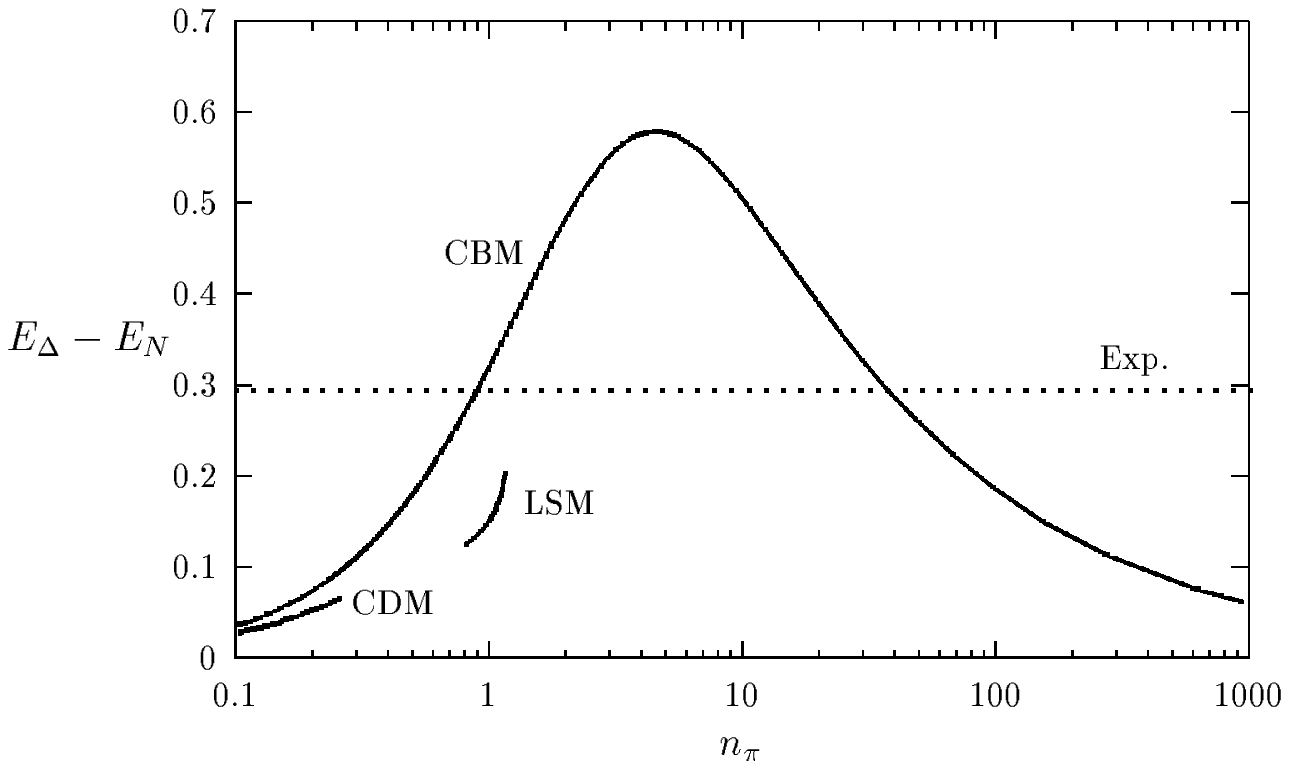}$$
\noindent Figure 2: $\Delta-N$ mass difference (in GeV) in CDM, LSM and CBM
as a function of $n_\pi$.}

\goodbreak
\noindent
{\bf Charge radii}
\smallskip\noindent
The charge radius squared of the proton is a weighted mean between
$\bra r^2_q\ket $
of the quark distribution and $\bra r^2_{\pi}\ket $ of the pionic
distribution. One might expect that $\bra r^2\ket $ increases with pion number
since the pionic distribution is broader. One is, however, surprised to find
that $\bra r^2\ket $ decreases with \np\ rather than increases (Fig.~3a).
The explanation is twofold.
First, the pionic distribution is not much broader. One intuitively thinks
of the extension of the Yukawa field
which is of the order of $1/m_\pi \sim $ 1.4 fm.
The charge distribution of pions, however, is much narrower than
the field.  This can be seen from the relation between the pion field
$\Phi$ and the ``pion wavefunction" $\phi$ in momentum space
$$\tilde{\Phi}(k)\sim (\tilde{\phi}^*(k)+\tilde{\phi}(k))/\sqrt{2\omega_k}\>,
\qquad \omega_k\approx k\>. $$
Since $\tilde{\phi}(k) \sim \sqrt{k}\, \tilde{\Phi}(k)$
is enhanced at large momenta, the
charge distribution in coordinate space is enhanced at smaller $r$.
For example, for typical model parameters in the LSM ($g=4.8$)
we get for the pion profile a radius squared 0.87 fm$^2$,
not drastically larger than for the quark profile (0.48~fm$^2$).

On the other hand, all radial profiles (of quarks and pions)
shrink with increasing \np\ (increasing interaction strength $g$)
which overcompensates the increase due to the higher weight of pions.

The charge radius squared of the neutron (lower curves in Fig.~3a)
remains small
and negative for all values of \np. The reason is similar as before,
the negative charge distribution of pions is not much broader than
the positive charge distribution of the quark core so that they
almost cancel, consistently with the experimental value.

\bigskip
\goodbreak
\medskip
\noindent
{\bf Magnetic moments}
\smallskip\noindent
While the {\it isovector magnetic moment} where quarks and pions
contribute is well described the {\it isoscalar magnetic moment}
which lacks a pionic contribution is twice too small (Fig.~3b).
This is a common
feature of all models with quarks and pions (contrary
to models with quarks only which automatically give the correct ratio 5:1).
One should not conclude from this that the ``best" number of pions
is zero but rather that some additional degree of freedom is missing
on which the isoscalar magnetic moment is sensitive.

{\epsfxsize13.7cm$$\epsfbox[50 570 560 740]{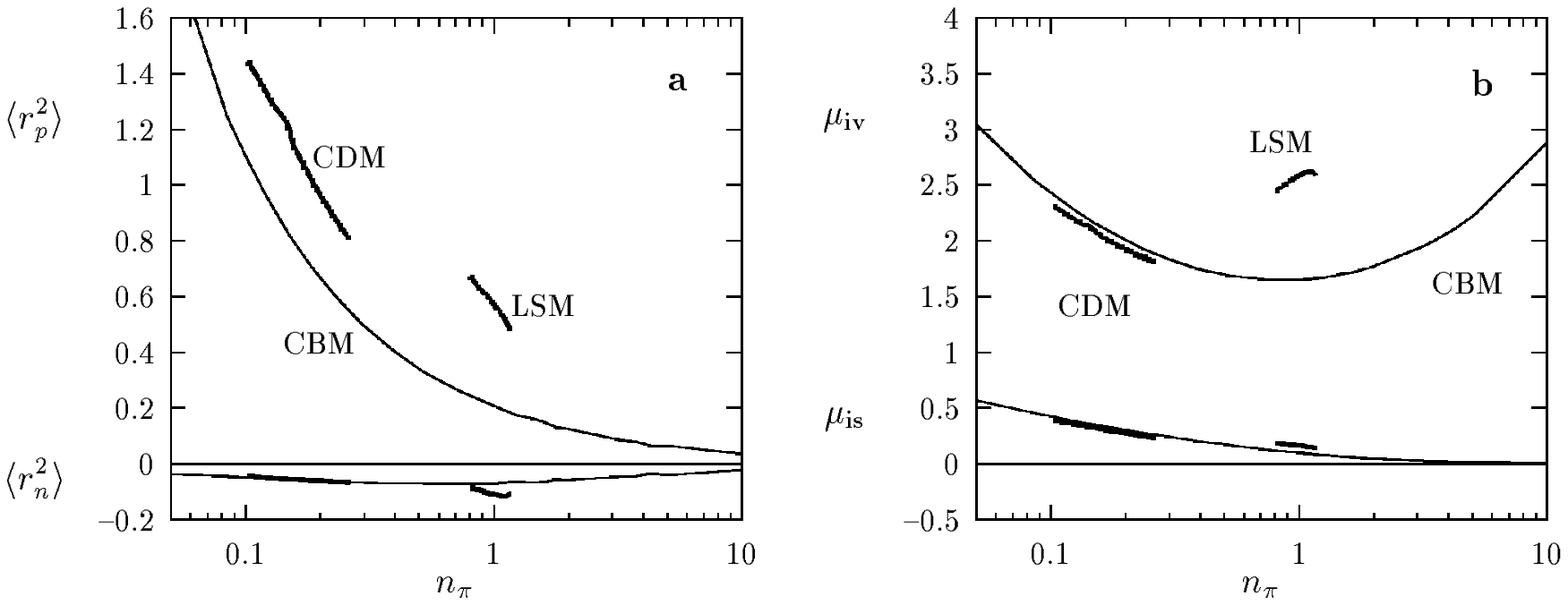}$$
\noindent Figure 3: Proton and neutron charge square radii [fm$^2$] ({\bf a})
and isoscalar and isovector magnetic moments
[n.m.] ({\bf b}) as a function of $n_\pi$ for CDM, LSM and CBM.}

\vfill\eject

\noindent
{\bf Axial charge}
\smallskip\noindent
In the LSM and the CDM, the axial charge appears as a cross term between the
pion and sigma fields. It is so large in the LSM (Fig.~4a)
because of the strong sigma field along with a relative large \np.
Therefore one cannot blame only \np\ but rather the strong sigma field.
In the CBM the pions do not contribute directly to this observable.

\bigskip
\goodbreak
\medskip
\noindent
{\bf Pion-nucleon coupling constant}
\smallskip\noindent
In the expression (\v Cibej, Fiolhais, Golli and M. Rosina, 1992)
$${m_\pi\over 2M_{N}}g_{\pi NN} =
{m_\pi^3\over 2\sqrt{3}}\, {4\pi\over\sqrt{2\pi\omega}}
\int_0^\infty r^3\xi_0(r)\d r\; \bra (a_{00}+a^{\dagger}_{00})\ket $$
the expectation value of $(a+a^\dagger)$ ``counts pions" and
increases approximately as $\sqrt{n_\pi}$, while the radial integral
decreases (Fig.~4b).
Because of these effects it is difficult to deduce the
``experimental" value of \np \ from this observable.
We can, however, better understand the
stability of the result
with respect to the coupling strength.
In the CBM the $g_{\pi NN}$ comes out too small
since we are using the physical
value for $f_\pi$. In the perturbative calculations of the CBM $f_\pi$
was a fitting parameter adjusted to reproduce the experimental value of
$g_{\pi NN}$ (Th\'eberge, Miller and Thomas, 1982).

{\epsfxsize13.7cm$$\epsfbox[50 550 560 740]{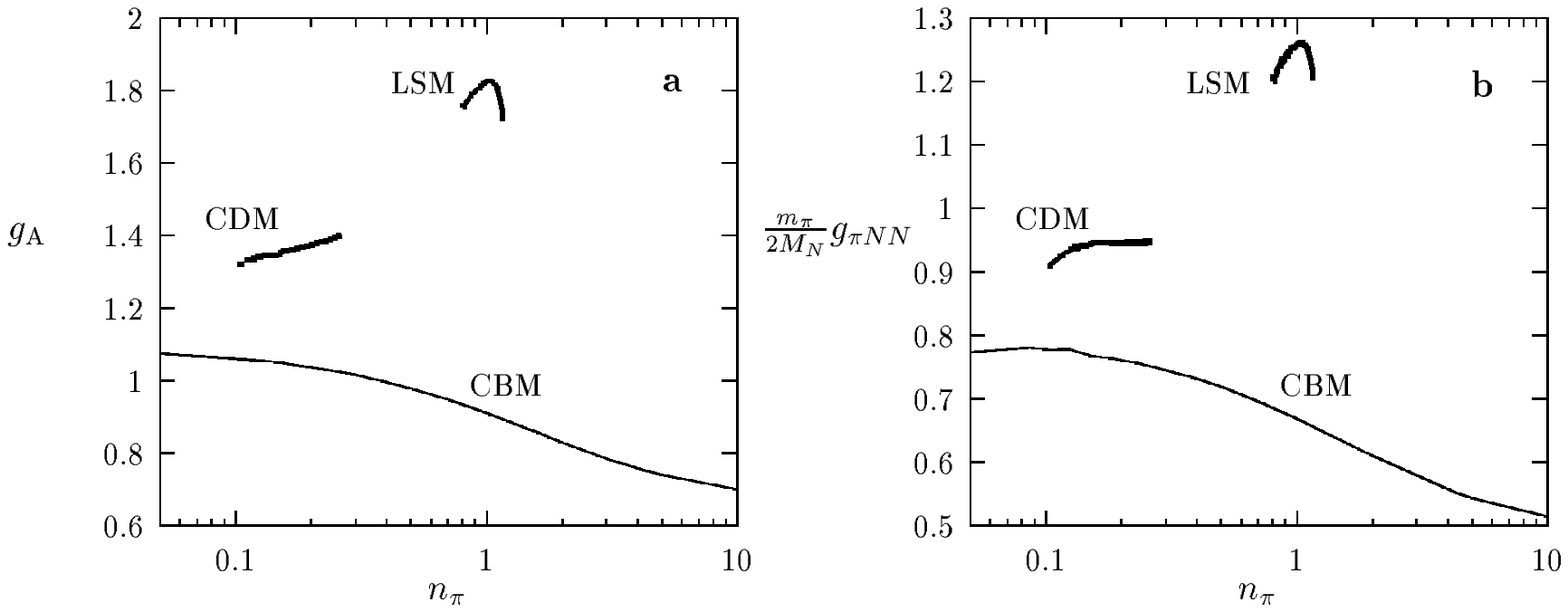}$$
\noindent Figure 4:
Axial coupling constant ({\bf a}) and the $\pi NN$ coupling constant ({\bf b})
as a function of $n_\pi$ for CDM, LSM and CBM.}

\bigskip
\goodbreak
\medskip
\noindent
{\bf Electric and magnetic polarizabilities }
\smallskip\noindent
The {\it electric polarizability} (Golli and Sraka, 1993)
comes twice too large in the LSM ($23\cdot~10^{-4}$~fm$^3$), particularly
due to the large seagull term
which possibly overcounts the pionic
contribution. The situation is opposite for the
{\it magnetic polarizability}: the relatively large \np\ around one which
we use is needed in order to compensate
the large paramagnetic term due to the virtual $\Delta$ excitation.
This virtual excitation cannot be avoided or diminished
(it contributes $20\cdot 10^{-4}$ fm$^3$), therefore the
diamagnetic effect of pions is necessary  to approach the observed
very small value (we get a pionic contribution $-10\cdot 10^{-4}$ fm$^3$).
A smaller number of pions which would improve the electric polarizability
would spoil the magnetic one. We argue that it is in the
electric polarizability that some effect is still missing and that a large
\np\ (which is needed for the magnetic polarizability) is not ruled out.
\bigskip
\medskip
\goodbreak
\noindent
\leftline{\hskip 2.8 truecm  {\bf ELECTROEXCITATION OF BARYON RESONANCES}}
\medskip
\nobreak\noindent

The amplitudes for electroproduction of low lying baryon resonances
can also yield sensible information on the pion content of the
nucleon and its excited states.
In particular, relatively large quadrupole $E2$ (or $E_{1+}$)
and $C2$ ($S_{1+}$) amplitudes in the vicinity of the $\Delta$(1232)
resonance can be almost entirely attributed to the cloud of
p-wave pions (Fiolhais, Golli and \v Sirca, 1995).
In quark models without the pion cloud, an unrealistically
strong admixture of d-state quarks would have to be introduced
in order to explain the experimental values.

In photoproduction, because of relatively low photon momentum
which, in the centre-of-mass frame, is given by
$k = k_W = (M^2_\Delta-M^2_N)/2M_\Delta = 258$~MeV,
the quadrupole photon only ``sees'' the tail of the pion cloud
in the region above 1~fm
whose behaviour is determined
by the Yukawa form and the $\pi N$ coupling constant.
If this mechanism is correct we can expect that all
those models containing the pion cloud which reproduce
$g_{\pi NN}$ are also able to reproduce well the
quadrupole amplitudes.
The number of pions is therefore not important here.
In order to be able to investigate the pion cloud in the
interior of the baryon we have to consider
electroproduction.
Since such a process involves virtual photons with
nonzero $Q^2$, $-Q^2=\omega^2-k^2$, it is possible to
``sit'' on the resonance while changing the photon momentum,
$k^2=[(M_\Delta^2+M_N^2+Q^2)/ 2M_\Delta]^2-
M_N^2$.
Because of kinematics, the $E2$ amplitude is very small
and difficult to measure precisely; on the other hand,
the scalar quadrupole amplitude $C2$
is clearly nonzero in electroproduction and may be an
important indication of the presence of a strong pion
cloud in the interior of the nucleon and the $\Delta$.
In Fig.~5 we plot the $C2$ amplitude calculated
as a function of $n_\pi$ in the three models
for photoproduction ($Q^2=0$) and for electroproduction at $Q^2=1$~GeV$^2$.
We see, as anticipated, that for the former its value only weakly
depends on the number of photons.
For $Q^2=1$~GeV$^2$, however, it strongly depends on
$n_\pi$; the experimental value
$C2=-11.8\pm 2.3\,\cdot 10^{-3} \,\hbox{GeV}^{-{1/2}}$
(Alder, 1972; Albrecht, 1971)
suggests a rather strong pion cloud.
(The experimental value at $Q^2=0$ is not measured directly.
Using current conservation and the Siegert
theorem it can be related to the value of $E2$, $E2=-4.4\pm 1.2\,
\cdot 10^{-3} \,\hbox{GeV}^{-{1/2}}$ (Particle Data Group, 1994).)

{\epsfxsize9.125cm$$\epsfbox[90 470 470 700]{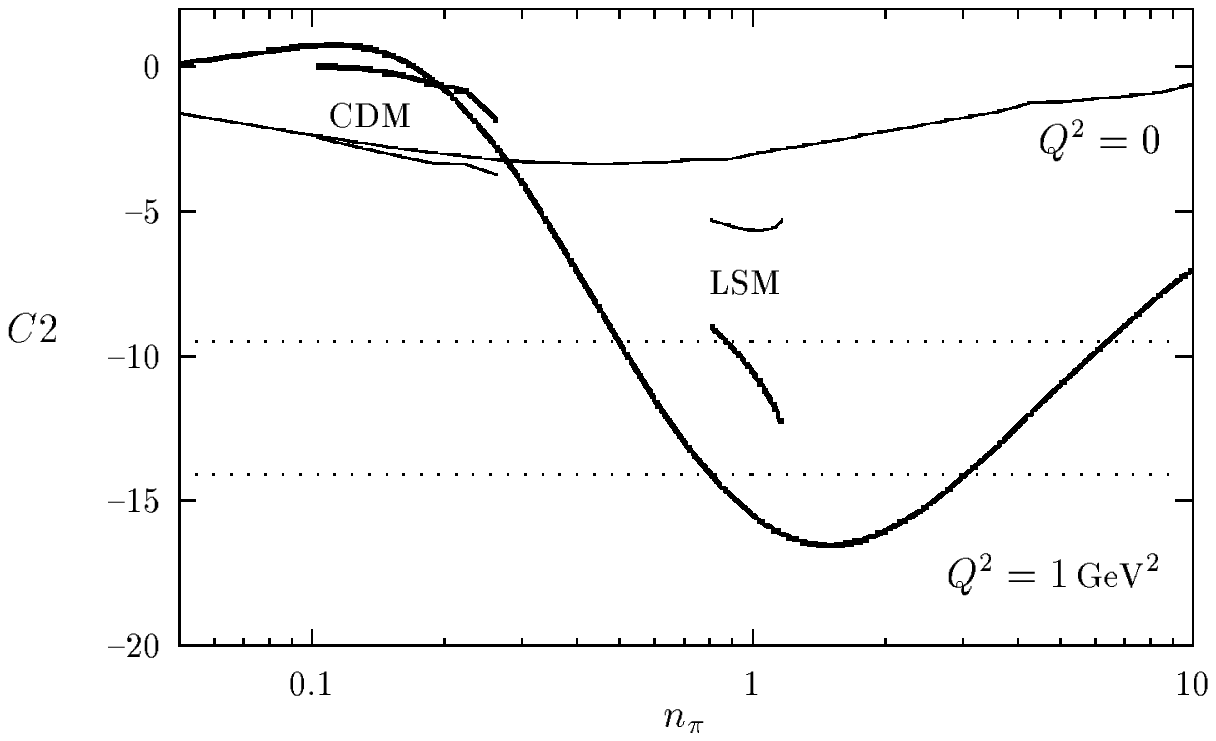}$$
\noindent Figure 5: Electroexcitation electric quadrupole amplitude $C2$
in units of $10^{-3}$GeV$^{-1/2}$ as a function of $n_\pi$ for CBM, LSM and CBM
at two values of photon virtuality: $Q^2=0\,$GeV (thin lines) and
$Q^2=1\,$GeV (thick lines). Dotted lines denote the experimental uncertainty
for $C2$ at $Q^2=1\,$GeV.}

\vfill\eject

\bigskip
\medskip
\goodbreak
\noindent
\leftline{\hskip 2.8 truecm  {\bf ELECTROPRODUCTION OF PIONS}}
\medskip
\nobreak\noindent

It is interesting to connect the number of ``theoretical pions"
to the number of pion-like partons
kicked out in the reactions $e+p \to e'+n+\pi^+$ or
$e+p \to e'+X+\pi^+$. Agreement would confirm the parton
picture of pions in the nucleon (in an impulse approximation).

A pioneering analysis of the pion distribution function in the nucleon
$G_{\pi^*/p}(x)$ was performed by G\"uttner, Chanfray, Pirner and Povh, 1984,
assuming
$$ {\d^2\sigma_L(ep\to e'X\pi)\over \d x\d Q^2} =G_{\pi^*/p}(x)
   {\d\sigma_{\hbox{\script el}}(e\pi^*\to e'\pi)\over \d Q^2} .$$
In principle, the pion distribution function $G$ is obtained
as a ratio between the measured pion electroproduction cross section
and the calculated cross section for the elastic
scattering of electron on pion. Integrating $G$, one obtains
the ``experimental number of pions" $n_\pi^{\hbox{\script exp}}$.
The authors derived the probability 3\% of the $\pi^+ n$
configuration in the proton and estimate further 1.5\%
for the $\pi^0 p$ and 3\% for the $\pi \Delta $  configurations.
This amounts at most to $n_\pi^{\hbox{\script exp}} = 0.08$,
favouring models with small number of pions.

This estimate is, however, inconclusive since
the DESY experiment  $e+p \to e'+n+\pi^+$
was not presented with full kinematics. The cross sections were given
as a function of
the squared momentum transfer
onto the target nucleon, $t = (p_{\gamma^*} - p_\pi)^2$ rather than
as a  function of the relevant
Bjorken variable $x$. Therefore, $\d^2\sigma (x,Q^2)$ was
approximately reconstructed using statistical assumptions.
Also, the proton and $\Delta$ final states were not measured in
this reaction and the corresponding contributions were only estimated.
Nevertheless, this analysis gave a strong stimulation for future research.
A repeated experiment and new experiments would be very important.

A new proposal
for the inclusive $e+p \to e'+X+\pi^+$ experiment has been presented
and a theoretical estimate \np\ = 0.16 given if all final states $X$,
also beyond nucleon and $\Delta$, are taken (Pirner and Povh, 1993).

In order to plan well the future experiments, different model predictions
have to be confronted so that one can anticipate the probable range of \np.
Recently, Baumg\"artner, Pirner, K\"onigsmann and Povh, 1995,
proposed a very interesting model in which most pions act as private
partons of each constituent quark. They could be probed
at low $x$ and very high energy of electrons (at HERA).
Since the model has been presented
by Pirner at this School (Pirner, 1995) we shall only sketch
the main idea.

If one assumes that the constituent quark consists of a bare quark
plus a bare quark with a pion cloud, the angular momentum and charge
gets distributed between the bare quark and the pion in a particular manner.
This can simultaneously improve the  Gottfried sum rule,
the integrated polarized structure function of the proton and  neutron,
the quark contribution to nucleon helicity and the axial charge.
The estimated pion admixture is 36\% per constituent quark
(around one pion per nucleon).

Such large \np\  resembles our results in the LSM. Of course, we do not
distinguish whether pions are correlated with individual quarks or with
the nucleon as a whole. We use a ``mean source approximation" for the
pions. Because the models are different the comparison is
inconclusive, but it is suggestive.

\vfill\eject

\noindent
\leftline{\hskip 2.8 truecm  {\bf CONCLUSION} } \medskip
\smallskip
\nobreak
\noindent

The contribution of the pion cloud to nucleon observables
can be conveniently analyzed
as a product of the number of pions \np\ times the specific contribution
per pion. While \np\ increases with increasing coupling strength
of pions to the quark core, the specific contribution may increase,
decrease or remain constant. Therefore we can classify several
typical cases.

(i) The electric charge radius of the proton,
the isovector magnetic moment, the axial coupling constant $g_A$ and the
$\pi NN$ coupling constant show
{\it indifference or a slow decrease with} \np.
The reason is that the specific contribution per pion decreases with
\np\ since all radial profiles shrink. Therefore these observables are
not suitable for ``counting pions".

(ii) The neutron electric charge radius squared contains a
{\it delicate cancellation} and cannot suggest a reliable value of \np.

(iii) The diamagnetic contribution to the magnetic polarizability
as well as the electroexcitation  of the $\Delta$ resonance show
a {\it strong dependence on} \np. They show that the contribution
of pions is essential and suggest \np\ $\sim 1$. Both theoretical
and experimental analysis are still in a preliminary stage but they
strongly encourage further theoretical and experimental studies of these
quantities which are so promising for pion counting.

(iv) The electric polarizability cannot come out well with same parameters
as the magnetic polarizability. The isoscalar magnetic moment does
not contain a pion contribution and is not
consistent with the isovector magnetic moment. The pion
contribution to the axial coupling constant
depends on the product of the pion and sigma fields
and not on  the pion field alone. Such {\it inconsistent or ambiguous cases}
are also not suitable to determine \np; other degrees of freedom are
required.

Different models predict  a wide range of values of \np.
It remains  a challenge to relate this theoretical concept to the number of
pions that are probed in electron scattering on the proton.

\bigskip
\leftline{\hskip 2.8 truecm  {\bf ACKNOWLEDGEMENT} } \medskip
\noindent

The research work presented in this report was partially supported by the
Ministry of Science and Technology of Slovenia, by the Calouste Gulbenkian
Foundation (Lisbon), and by the European Commission (contract
ERB-CIPA-CT-92-2287).
\bigskip
\leftline{\hskip 2.8 truecm  {\bf REFERENCES} } \medskip

\hang\noindent{J. Albrecht {\it et al.}, (1971).
{\sl Nucl. Phys. B} {\bf 25}, 1.}

\hang\noindent{J. C. Alder {\it et al.}, (1971).
{\sl Nucl. Phys. B} {\bf 46}, 573.}

\hang\noindent{L. Amoreira, M. Fiolhais, B. Golli and M. Rosina, (1995).
{\sl J. Phys. G}, accepted.}

\hang\noindent{S. Baumg\"artner, H. J. Pirner, K. K\"onigsmann
and B. Povh, (1995). Hep-ph/9507309 preprint.}

\hang\noindent{M. C. Birse, (1990). {\sl Prog. Part. Nucl. Phys.} {\bf 25}, 1.}

\hang\noindent{M. \v Cibej, M. Fiolhais, B. Golli and M. Rosina, (1992).
{\sl J. Phys. G} {\bf 18}, 49.}

\hang\noindent{M. Fiolhais, B. Golli, and
S. \v Sirca, (1995). Submitted to {\sl Phys. Lett. B.} }

\hang\noindent{B. Golli and R. Sraka, (1993).
{\sl Phys. Lett. B} {\bf 312}, 24.}

\hang\noindent{F. G\"uttner, G. Chanfray, H. J. Pirner and B. Povh,
(1984). {\sl Nucl. Phys. A} {\bf 429}, 727.}

\hang\noindent{Particle Data Group, (1994).
{\sl Phys. Rev. D} {\bf 50}, 1712.}

\hang\noindent{H. J. Pirner, (1995).
{ Structure of the constituent quark}, {\sl these Proceedings}.}

\hang\noindent{H. J. Pirner and B. Povh, (1993).
{\sl Proc. Conf. ``The ELFE Project: an Electron
 Laboratory for Europe'', Mainz, Germany, 7--9 Oct. 1992},
{ed. J. Arvieux}, {\it Ital. Phys. Soc., Bologna}, 45.}

\hang\noindent{M. Rosina, (1994).
{\sl Proc. Int. Conf. on Many-Body Physics, Coimbra, Portugal,
20--25~Sept.~1993}, {eds. C. Fiolhais, M. Fiolhais, C. Sousa and
J. N. Urbano,} {\it World Scientific, Singapore}, 161. }

\hang\noindent{S. Th\'eberge, G. A. Miller and A. W. Thomas, (1982).
{\sl Can. J. Phys.} {\bf 60}, 59.}

\hang\noindent{A. W. Thomas, (1983). {\sl Adv. Nucl. Phys.}
{\bf 13}, 1.}

\vfill
\eject
\bye